\documentclass[12pt]{article}
\usepackage{epsfig,eurosym}
\begin{document}
\title{Extension of the Uhlenbeck-Ford Model with an Attraction}
\date{\today} 
\author{J. M. J. van Leeuwen}
\maketitle
\begin{center}
Instituut-Lorentz, Universiteit Leiden,\\
Niels Bohrweg 2, 2333 CA Leiden, The Netherlands.
\end{center}

\begin{abstract}
  The Uhlenbeck-Ford model for soft repulsion, which has only a
  repulsive interaction, is extended by inclusion of an attraction.
  This extension still allows an analytical evaluation of the
  virial coefficients. The integrals over the graph contributions
  are reduced to a combinatorial problem.
  We have calculated the virial coefficients to order 6 in the density.
  A link is made between this model and more common
  interactions, like the 12-6 Lennard-Jones potential.
  
\end{abstract}

\section{Introduction}
The expansion of the thermodynamic properties in terms of a power
series in the density is the oldest tool in the systematic study of
the properties of dense gases. It has the form \cite{mayer}
\begin{equation} \label{a1}
  \frac{p}{k_B T} = \sum_{l=1} B_l\, n^l,
\end{equation}
where $p$ is the pressure, $T$ the absolute temperature and $n$ the
number density. The coefficients $B_l$ are called the virial
coefficients ($B_1 = 1$). The expansion is unique in the sense
that each term in the series has an explicit prescription for its
calculation. Mayer \cite{mayer, montroll} was the first to introduce a
graphical representation for the various contributions.
For low densities a few terms suffice and in this
domain the virial expansion has proven to be an invaluable tool
for the computation of the thermodynamic properties. Not only the
pressure but also the other thermodynamic quantities can be expanded
in a series in the density $n$ expressable in the same coefficients $B_l$.
A similar expansion for the transport properties has been attempted
\cite{cohen, sengers, dorfman, vanl}, but even the first correction
to ideal gas behavior
leads to divergencies and it turns out that the transport
properties are not expandable in a power series in the density.

The virial coefficients $B_l$ are found as $l$-fold integrals
\begin{equation} \label{a2}
  B_l = -\frac{l-1}{V \, l!} \int d {\bf r_1}  d {\bf r_2}  \cdots
  d {\bf r_l} \, H_l ({\bf r_1}, \cdots , {\bf r_l}),
\end{equation}
where the function $H_l$ is represented by graphs with $l$ vertices and
a number of occupied edges, each carrying the (Mayer) function
\begin{equation} \label{a3}
  f(r_{ij}) = \exp [-V(r_{ij})/(k_B T)] - 1
\end{equation}
with $V(r_{ij})$ the intermolecular potential between the particles $i$
and $j$.

The problem of evaluating the virial coefficients is twofold.
The first part is the generation of the graphs and the second part
are the integrals associated with the graphs.
The generation of the graphs is general and independent
of the interaction potential of the
gas, while the evaluation of the integrals is strongly dependent on the
interaction. Usually finding the graphs is not the limiting factor
since the integrals become already for order 5 a too-complicated
for e.g. a Lennard-Jones interaction.
An exception forms the so-called hard-sphere interaction where the
virial coefficients are evaluated up to order 10. 
The hard-sphere model has been thoroughly investigated in all
dimensions \cite{clisby1,clisby2,clisby3,clisby4}, in
particular in its relation to the convergence of the virial series.

The generation of graphs has been extensively studied by the
mathematicians, who have developed an efficient algorithm for this
problem. Even with their efficiency the enumeration gets laborious
around order 10, due to the more than
exponential growth of the number of relevant graphs. For our purpose
the bare generation of the graphs is not sufficient, we have to know
also the symmetry properties of the graph. As this is a job
valid for all systems, one has to generate the sequence only once
and the time spend on it is generally not the bottle-neck for the
calculation. By straightforwardly generating all graphs and
eliminating those that do not qualify, we could reach all the graphs
up to $l=9$ vertices together with their symmetry properties.

It is an old idea, due to Uhlenbeck and Ford \cite{uhlenbeck},
to simplify the evaluation of the graphs by replacing
the Mayer function $f(r)$ by a gaussian
\begin{equation} \label{a4}
  f(r)  \Rightarrow - \exp \left(-\frac{r^2}{2 a^2}\right).
\end{equation}
One can view this replacement as an approximation of a
potential, but also as a model on its own. 
As the corresponding potential is positive everywhere,
with an infinite limit for  $r=0$,
the Uhlenbeck-Ford (UF) model can be considered as representing a soft
repulsive system with range $a$. The advantage of this
model is that the integration of any graph contribution becomes a
gaussian integral, which can be evaluated analytically.
The model Eq.~(\ref{a4}) contains only
one parameter, the length scale $a$ which can be combined with the
density $n$ to a dimensionless measure for the density.

Recently this idea has been vigorously picked up by Leite et al. 
\cite{leite, leite1}. They evaluated not only the virial
coefficients up to $B_{13}$ (!), but they also investigated
the model extensively with molecular dynamics. In addition
they considered the scaled Uhlenbeck-Ford model which has the
potential multiplied by an integer. That makes the core more repulsive
while keeping the graph contributions still integrable.
They also gave a survey of the properties and
applications of the UF model.

In this paper we make an extension of the
model by adding an attractive part to the potential.
One could do this by taking $f(r)$ as a sum of two gaussians
\begin{equation} \label{a5}
  f(r) = -(1+A) \exp \left(-\frac{r^2}{2 a^2_1}\right)+
  A \left(-\frac{r^2}{2 a^2_2}\right)
\end{equation}
Since $f(0)=-1$ the potential has again a repulsive core.
By taking the range $a_2 > a_1$ the potential gets
an attractive tail (for $A>0$).
By playing with the parameters one can influence the range and the
depth of the attractive well in the potential. The price to be paid
is considerably more computational effort and each choice of the
parameters requires the full evaluation of the virial coefficients.
A more severe limitation comes from the fact that we want to have the
range $a_2$ of the attraction not too different from the range $a_1$
of the repulsion. As the two terms in Eq.~(\ref{a3}) have opposite
signs, the contributions tend to cancel, making the sum much smaller
than the individual terms, which easily leads to numerical
errors. This is a delicate problem which generally plagues the 
evaluation of the virial coefficients, since they are the result of
many contributions with uncorrelated signs.

A remedy for this danger is to see the difference as a derivative,
leading to the Mayer function
\begin{equation} \label{a6}
  f(r) = (-1+ A (r/a)^ 2) \exp(-(r/a)^2/2).
\end{equation}
This paper is devoted to the evaluation of the virial coefficients
resulting from this Mayer function. They become functions
$B_l(A)$ in the form of a finite polynomial
\begin{equation} \label{a10}
  B_l (A) =\sum_{0 \leq k \leq m} B_{l,k}  \, A^k.
\end{equation}
The $k$-th power of $A$ comes from graphs with at least $k$
edges. The maximum power of $A$ comes from the graph with all
$m$ edges occupied with
\begin{equation} \label{a11}
  m=l(l-1)/2,
\end{equation}
So the expansion in powers of $A$ terminates at the $m$-th power.
For $A=0$ only the $k=0$ terms contribute and the series 
Eq.~(\ref{a10}) reduces to that for the soft repulsive potential. 

The model, described by Eq.~(\ref{a6}), contains only one free
parameter $A$ apart from the scale parameter $\alpha$. In that sense
it is less versatile than the model given by Eq.~(\ref{a5}), which
contains three extra parameters. But the result as presented in
Eq.~(\ref{a10}) is more useful. It gives explicitly the $A$
dependence of the virial coefficients $B_l(A)$,
through the $B_{l,k}$ as a finite series in powers of $A$.

There is another advantage of the extension Eq.~(\ref{a6}).
One of the drawbacks of the Uhlenbeck-Ford model is
that there is no temperature dependence, like in the hard sphere
model. Whereas in the hard sphere model this is a consequence of the
fact that the potential is either zero or infinite, the
potential has in the
Uhlenbeck-Ford model has a distance dependent structure,
which cannot be varied by a continuous amplitude.
With the choice Eq.~(\ref{a6}) the thermodynamic properties
become functions of the density $n$ and the parameter $A$, which
mimics the variable $1/T$. A large $A$ means a deep well in the
reduced potential $V(r)/(k_B T)$ or a low $T$ at fixed $V(r)$.

It is customary in this field to present the results in a
dimensionless way, for which we use the second virial coefficient
$B_{2,0}$ of the repulsive part 
\begin{equation} \label{a12}
  B_{2,0} =\frac{ (a \sqrt{2 \pi} )^d}{2} \equiv v_0,
\end{equation}
with $d$ the dimension of the system. With this molecular volume
we construct the dimensionless density $n^*$ as
\begin{equation} \label{a13}
  n^* =  n v_0
\end{equation}
and turn the virial series into
\begin{equation} \label{a14}
  \frac{p \, v_0}{k_B T} = n^*+\sum_{l \geq 2}\, \sum_{0 \leq k \leq m} B^*_{l,k}
  \, A^k \, n^{*\,l}.
\end{equation}
with
\begin{equation} \label{a15}
  B_{l,k}^* = \frac{B_{l,k}}{v_0^{l-1}}.
\end{equation}
The coefficients $B^*_{l,k} $ are dimensionless numbers. They
constitute the objects to be calculated in this paper.

\section{Properties of the potential and the parameters}

The potential $V(r)$ corresponding to the Mayer function $f(r)$,
given by Eq.~(\ref{a6}), reads
\begin{equation} \label{a0}
  \frac{V(r)}{k_B T} = - \log \left[1 + \left(-1+A \frac{r^2}{a^2} \right)
    \exp \left(-\frac{r^2}{2a^2} \right) \right]
\end{equation}
\begin{figure}[h]
\begin{center}
  \epsfxsize=0.7\linewidth  \epsffile{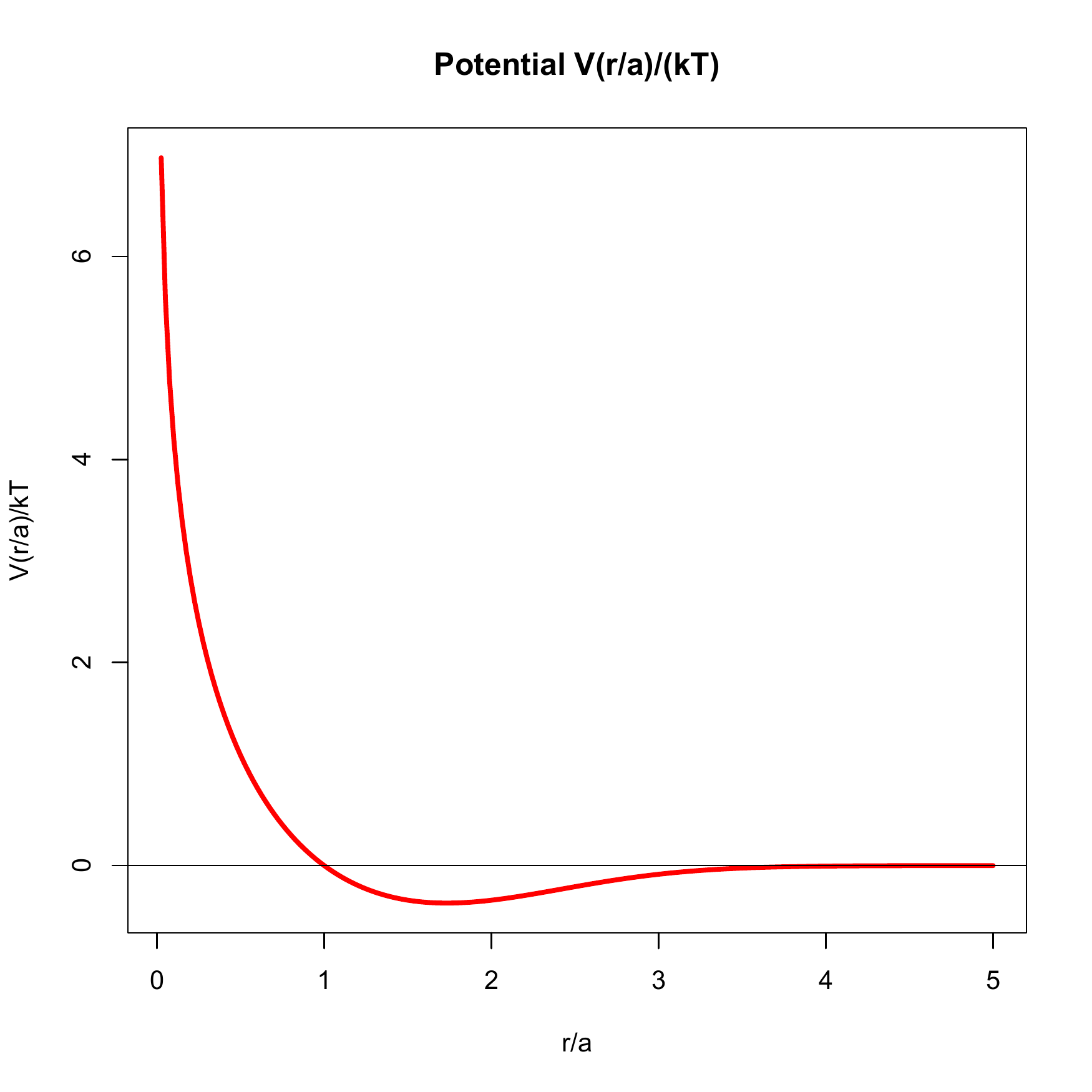}
  \vspace*{-0.1cm}
  
  \caption{The potential of the extended Uhlenbeck-Ford model for $A=1$}
  \label{potential}
  \end{center}
\end{figure}
In Fig.~\ref{potential} we show its behavior for the value $A=1$.
The potential  is zero at the distance $r=r_0$, the point
where $f(r)=0$
\begin{equation} \label{a7}
  r_0= a/(A)^{1/2}.
\end{equation}
For distances $r$ smaller $r_0$ the particles feel the repulsive
part of the potential and beyond $r_0$ they are in the attractive
well of the potential.
The potential $V(r)$ has a minimum where $f(r)$ has a maximum, which
occurs at the position $r=r_m$
\begin{equation} \label{a8}
  r_m =a \left(\frac{1+2 A}{A} \right)^{1/2} = r_0 (1+2 A)^{1/2}.
\end{equation} 
The value of $f(r_m)$ and the depth of the potential $V(r_m)$ are
related as
\begin{equation} \label{a9}
  \frac{ V(r_m)}{k_B T} = -\log (1 + f(r_m)) =
  -\log \left[ 1 + 2A \exp-\left(\frac{1+2A}{2A} \right)\right].
\end{equation}

In order to illustrate the influence of the parameters $n^*$ and
$A$, we consider the configuration where the particles occupy the
points of an fcc-lattice. This a close-packed arrangement where
the distance of any particle to its nearest neighbors equals the
same value $b$, given by the density $n$ as
\begin{equation} \label{g1}
  n = \frac{4}{(b \sqrt{2})^3}, \quad \quad {\rm or}
  \quad \quad b = \left(\frac {\sqrt{2}}{n }\right)^{1/3}.
\end{equation}
We compare this distance with the distance $r_0$, given by
Eq.~(\ref{a7}), using the reduced density $n^*$ as defined in
Eq.~(\ref{a12})
\begin{equation} \label{g2}
  b = \left(\frac{v_0 \sqrt{2} }{n^* } \right)^{1/3} =
  \left( \frac{(a \sqrt{2 \pi})^3} {n^* \sqrt{2}} \right)^{1/3}=
  r_0  \frac{2^{1/3}\sqrt{\pi A}}{(n^*)^{1/3}}
\end{equation}
Small $A$ and/or large $n^*$ force the fcc distance to be smaller than $r_0$
and the particles are in the repulsive part of the potential.
For larger $A$ or smaller $n^*$ they are in the attractive well of
the potential.

For potentials with a finite hard core radius $\sigma$, the system
forms a solid {\it before} the distance $b$ equals $\sigma$.
Larger densities $n$ are not possible as they lead to an infinite
pressure. Likely this geometric ordering is absent in the
Uhlenbeck-Ford model. Also the behavior of the virial coefficients
supports this idea as there is no indication of a diverging virial
series for a finite radius of convergence.

\section{The Gaussian Integrals} \label{gauss}

\begin{figure}[h]
\begin{center}
  \epsfxsize=\linewidth  \epsffile{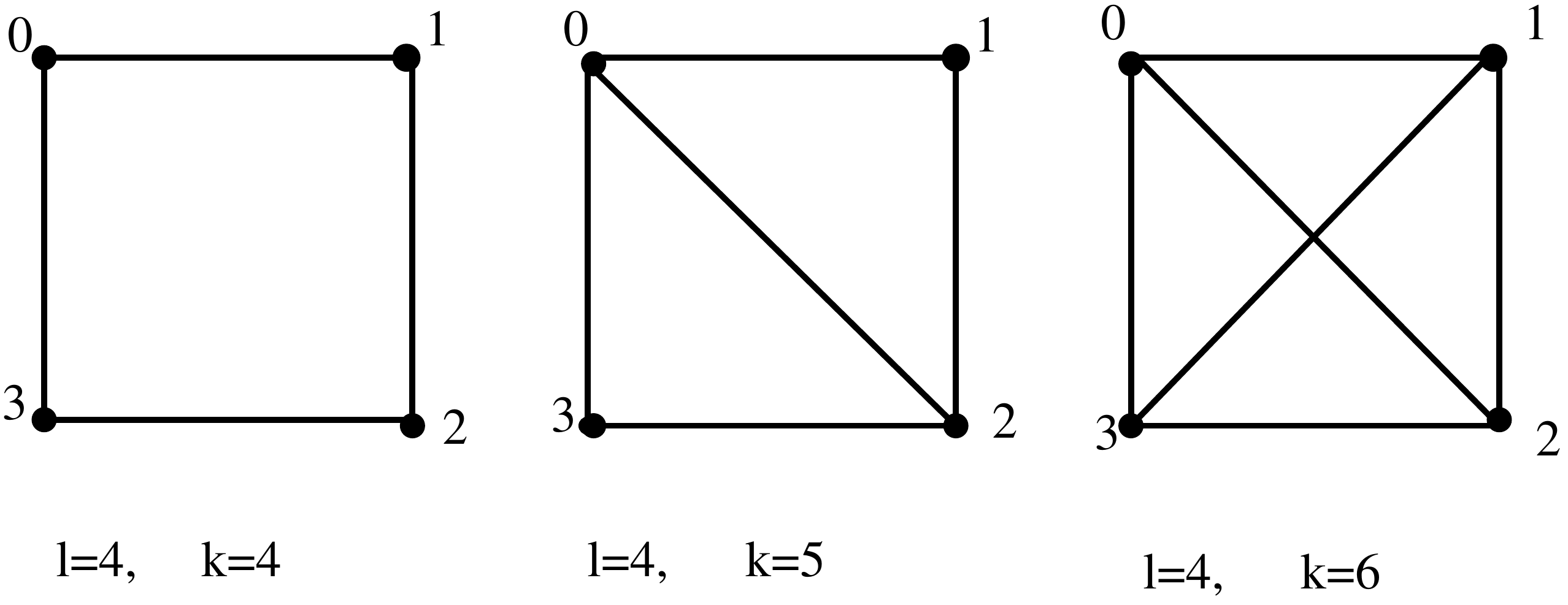}
  \vspace*{-0.5cm}
  
  \caption{The graphs for the fourth virial coefficient $B_4$} 
  \label{graphs}
  \end{center}
\end{figure}

In this Section we illustrate the evaluation of the virial
coefficients. In Fig. \ref{graphs} we show the three graphs
contributing to the fourth virial coefficient. The graphs have to be
doubly-connected, i.e. they remain connected if one of the vertices
is removed \cite{montroll}. We rewrite the Mayer function Eq.~(\ref{a6}) as
\begin{equation} \label{b1} 
f(r)= -\left( 1 + 2 A
  \frac{\partial}{\partial \alpha} \right)
\exp \left(-\frac{\alpha r^2}{2 a^2}\right)
\end{equation}
and set later $\alpha=1$. So we have to evaluate the graph
contribution as a function of the $\alpha_{ij}$ on the edge $(ij)$.
As we shall see this yields a relatively simple polynomial in
the $\alpha_{ij}$.
Then perform the differentiations with respect to the $\alpha_{ij}$
and set them equal to $\alpha_{ij}=1$. 
We illustrate the handling of the
Gaussian integrals for the fourth virial coefficient. The integrals
are of the form
\begin{equation} \label{b2}
  C_4 = \int d {\bf r_0}  d {\bf r_1}  d {\bf r_2}  d {\bf r_3}
  \exp\left(- \sum_{(ij)} \frac{\alpha_{ij} r_{ij}^2}{2 a^2} \right),
\end{equation}
where the sum runs over the occupied edges $(ij)$.

In the integral over the vertices we may carry out a permutation of
the vertices without changing the contribution. In general such a
permutation leads to a different connectivity, i.e. to a different graph.
These graphs can be taken together. It amounts to a cancelation of the
$l!$ in the denominator of Eq.~(\ref{a2}), unless the graph has
symmetries. The first graph in Fig.~(\ref{graphs}) has 8 symmetry
operations: 4 cyclic permutations multiplied by the factor 2 of
the reflection symmetry. Thus the 4! permutations yield only
3 different graphs. The symmetry number is the size of the symmetry
group. For of the first graph it is 8,
that of the second graph is 4 and of the last graph it is 24. The
symmetry number replaces the $l!$ in the denominator of Eq.~(\ref{a2}).

The symmetries of a graph are important, not only because of the
symmetry number, but also in reducing the computational 
evaluation. Therefore it is useful to generate the symmetry
operations together with the generation of the graph.

The integrand of Eq.~(\ref{b2}) is translational invariant, which
is exploited by making the shift
\begin{equation} \label{b3}
  {\bf r}'_i = {\bf r}_i -{\bf r}_0,
\end{equation} 
for $i \neq 0$. As result ${\bf r_0}$ does not appear in the
integrand anymore and the integral over $\bf r_0$ yields a volume
factor in the integration in Eq.~(\ref{b2}). 
By the shift Eq.~(\ref{b3}), the exponent is changed into
\begin{equation} \label{b4}
  \sum_{(ij)} \alpha_{ij} \frac{ r_{ij}^2}{2 a^2} =
  \sum_{i=1}^3 \sum_{j=1}^3 w_{i,j} \, \frac{\bf r'_i \, r'_j}{2a^2}.
\end{equation}
The matrix $w_{i,j}$  is given by (only non-zero $i$ and $j$).
\begin{equation} \label{b5}
  w_{i,j} = -\alpha_{ij} \quad i \neq j, \quad \quad \quad
  w_{i,i} = \alpha_{0i}+ \sum_{j \neq i}\alpha_{ij}.
\end{equation}
After the elimination of the integration over ${\bf r}_0$ the
contribution $C_4$ becomes
\begin{equation} \label{b6}
  C_4 = V \int d {\bf r'_1}  d {\bf r'_2}  d {\bf r'_3}
  \exp \left(-\sum_{i=1}^3 \sum_{j=1}^3 w_{i,j} \, \frac {\bf r'_i \, r'_j}
  {2a^2}\right)
\end{equation}
The connectivity matrix $w_{i,j}$ is characteristic for the graph.

\begin{figure}[h]
\begin{center}
  \epsfxsize=0.7\linewidth  \epsffile{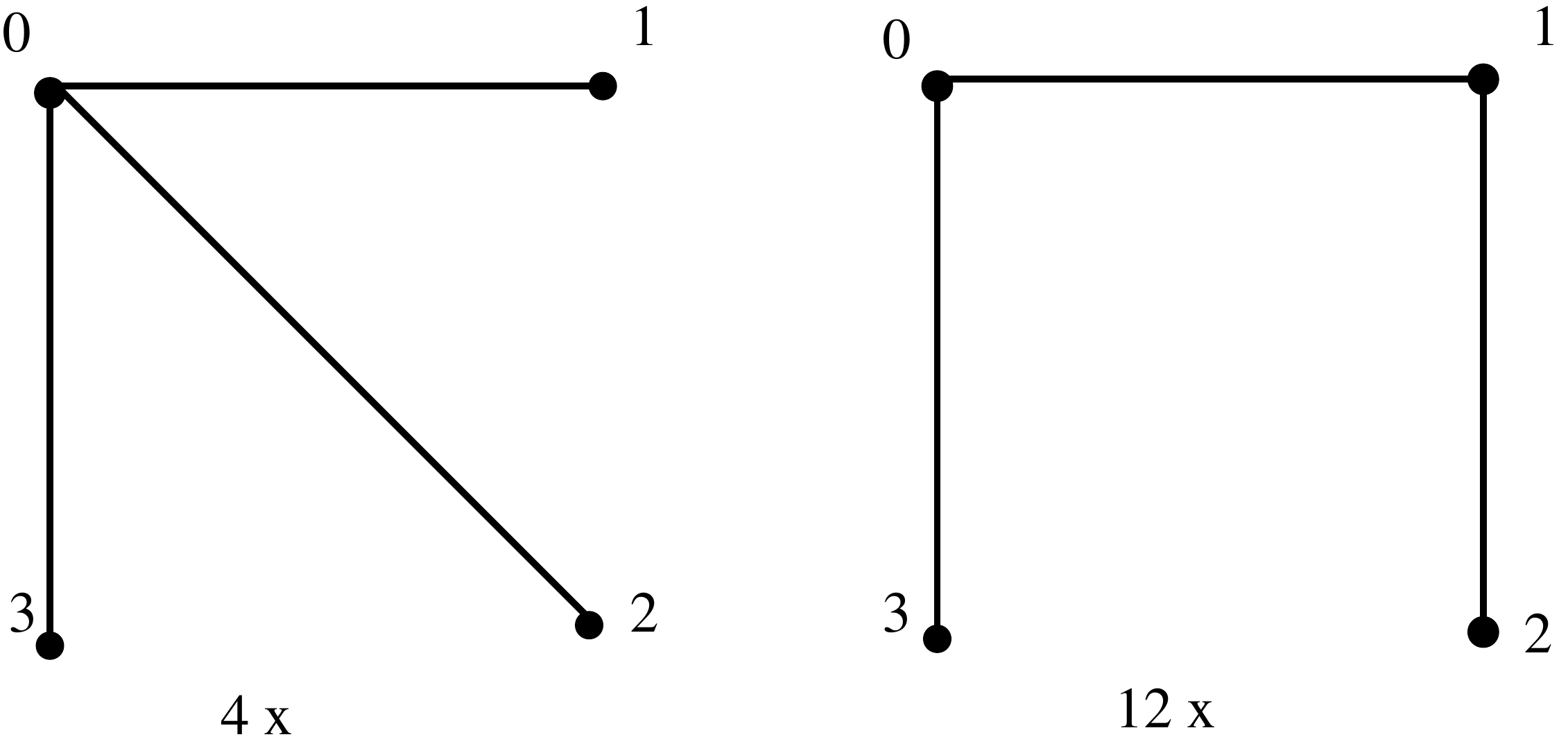}
  \vspace*{-0.1cm}
  
  \caption{The 16 spanning trees for fully occupied graph of $l=4$ vertices} 
  \label{spanning}
  \end{center}
\end{figure}

The gaussian integral is evaluated by diagonalization of
the matrix $w_{i,j}$, leading to the eigenvalues $\lambda_i$ on the
diagonal. Each eigenmode gives a factor $([2 \pi a^2]^3/\lambda_i)^{d/2}$
with $d$ the dimension of space. Since the product of the eigenvalues
equals the determinant we find for $C_4$
\begin{equation} \label{b7}
  C_4 = V \left(\frac{[2 \pi a^2]^3}{\det(w)} \right)^{d/2}.
\end{equation}
Written out the determinant reads for the graph with $k=4$
\begin{center}
  det($w$)\, = 
  \begin{tabular}{|ccc|}
      $\alpha_{01}$+$\alpha_{12} $ & -$\alpha_{12}$ &  0\\*[2mm]
      -$\alpha_{12}$ & $\alpha_{12}$+$\alpha_{23}$ & -$\alpha_{23}$ \\*[2mm]
      0 & -$\alpha_{23}$  & $\alpha_{03}$ + $\alpha_{23}$ \\*[2mm]
  \end{tabular}
\end{center}
Working out the terms of the determinant yields the function
\begin{equation} \label{b8}
  P(\alpha_{ij})=\alpha_{01} \alpha_{12} \alpha_{23} +
  \alpha_{12} \alpha_{23} \alpha_{03} +
  \alpha_{23} \alpha_{03} \alpha_{01} +
  \alpha_{03} \alpha_{01} \alpha_{12}.
\end{equation}
This function, which is characteristic for the graph, is called the
Kirchhoff polynomial.
Note that all terms have $l-1\, (=3)$ factors $\alpha_{ij}$ and
no higher powers of the $\alpha_{ij}$ occur.
If we draw an edge for every
$\alpha$ appearing in a term we get the four spanning trees of the
graph. In fact there is a general rule \cite{theorem} of Kirchhoff
linking the graph
function to the set of spanning trees of the graph
\begin{center}
{\it the Kirchhoff polynomial is given by the set of numbered
  spanning trees of graph.}
\end{center}
A spanning tree is a connected graph with the minimal number of
edges. So for a graph of $l$ vertices, a spanning tree has $l-1$
edges. The full set of spanning trees of 4 vertices are shown in
Fig.~\ref{spanning}. The 4 terms in Eq~(\ref{b8}) correspond to the
4 possible numberings of the spanning tree. The value of the
graph function for all $\alpha_{ij}=1$ is the number of spanning trees.

Instead of the determinant we use the Kirchhoff polynomial as it
contains all the information needed for the differentiations.
Spanning trees are easily generated and therefore a convenient way
to calculate the Krichhoff polynomial. For the fully occupied graph of
$l$ vertices the number of spanning trees is given by the Caley
theorem \cite{caley}
\begin{equation} \label{b9}
  P(\alpha_{ij}=1) = l^{l-2}. 
\end{equation}
The fully occupied graph has the maximum number of spanning trees.

The Kirchhoff polynomial of the fully occupied graph with $l$ vertices,
contains also all the information needed for calculation of the other
graphs. The Kirchhoff polynomial for graphs
with lesser occupied edges follows from the fully occupied graph
by setting the $\alpha_{ij}=0$ for the empty edges. 

Although the generation of the graph polynomial is simple, the
calculation of the derivatives is quite involved for the higher
derivatives. The easiest graph contribution is
the one without derivatives i.e. the value of $B^*_{l,0}$. Without
too much computational effort we obtained the following list of
$B^*_{l,0}$ (for $d=3$).
\begin{center}
  \begin{tabular}{|c|r|}
    \hline 
    & \\[-2mm]
    $B^*_{2,0} $ & 1.0000000000  \\[2mm]
    \hline
    & \\[-2mm]
    $B^*_{3.0} $ & 0.2566001196 \\[2mm]
    \hline
       & \\[-2mm]
    $B^*_{4,0} $ & -0.1254599571 \\[2mm]
        \hline
    & \\[-2mm]
    $B^*_{5,0} $ & 0.0133256552 \\[2mm]
        \hline
    & \\[-2mm]
    $B^*_{6,0} $ & 0.0384609358 \\[2mm]
    \hline
    & \\[-2mm]
    $B^*_{7,0} $ & -0.0330834429 \\[2mm]
        \hline
    & \\[-2mm]
    $B^*_{8,0} $ & 0.0041824181 \\[2mm]
        \hline
    & \\[-2mm]
  $B^*_{9,0} $ & 0.0151976158 \\[2mm]
    \hline
    
  \end{tabular}
  
\end{center}
\begin{center}
  Table 1.  Virial coefficients $B_{l,0}$ for pure repulsion.
\end{center}
These numbers agree to all digits with those given in \cite{leite}.
As they were obtained prior to noticing this paper,
it can be seen as a mutual confirmation.

\section{The Lower Derivatives}

In order to find the algorithm for the derivatives we first
concentrate on the low derivatives as occurring in the first few
virial coefficients. The second virial coefficient has a single
graph of two vertices connected by an edge. This is also the
spanning tree of the graph and the associated polynomial reads
\begin{equation} \label{c1}
  P(\alpha)=\alpha.
\end{equation}
For the virial coefficient we find
\begin{equation} \label{c2}
  B_2 = \frac{1}{2} \left(1 + 2 A \frac{\partial}{\partial \alpha}
  \right)  \left( \frac{2 \pi a^2}{P(\alpha)} \right)^{d/2}_{\alpha=1} =
   \frac{ [2 \pi a^2]^{d/2}}{2} (1 - d A).
\end{equation} 
Thus we have the two coefficients, $B_{2,0}$ given by
Eq.~(\ref{a12}) and 
\begin{equation} \label{c3}
  B_{2,1} = - d B_{2,0} \quad \quad {\rm or} \quad \quad
  B^*_{2,1} =-d.
\end{equation}

The third virial coefficient also involves a single graph, the
fully occupied graph with three vertices connected by three edges.
Numbering the edges 1,2 and 3
we obtain the graph polynomial
\begin{equation} \label{c4}
  P(\alpha_1, \alpha_2, \alpha_3) =  \alpha_1  \alpha_2+
  \alpha_2 \alpha_3 +  \alpha_3 \alpha_1.
\end{equation}
The third virial coefficient is then found as
\begin{equation} \label{c5}
  B_3=\frac{(2 \pi a^2)^d}{3}
  \left(1 + 2 A \frac{\partial}{\partial \alpha_1} \right)
  \left(1 + 2 A \frac{\partial}{\partial \alpha_2} \right)
  \left(1 + 2 A \frac{\partial}{\partial \alpha_3} \right)
  P (\alpha_1, \alpha_2, \alpha_3)^{-d/2}
\end{equation} 
Performing the differentiations and setting the $\alpha$'s equal to 1
yield for $B^*_3$ 
\begin{equation} \label{c6}
  B_3^*=\frac{4}{3^{1+d/2}} \left[1-2 d A + \frac{2 d (2d+1)}{3} A^2 -
  \frac{4 d(d+2)(2d-1)}{27} A^3\right],  
\end{equation}
from which the components $B^*_{3,k}$ follow as the coefficients
of the powers in $A$.

Generally, let $\kappa, \lambda, \mu$ be a choice of three edges
out of the $k$ edges of the graph. Then we have the relations
\begin{eqnarray} \label{c7}
   P^{d/2}\, \partial_\kappa\left( P^{-d/2} \right) & = & -\frac{d}{2P} \,
     \partial_\kappa P,  \\*[2mm]
   P^{d/2} \partial_{\kappa} \partial_{\lambda} \left( P^{-d/2} \right)  & = & -
     \frac{d}{2P} \, \partial_{\kappa} \partial_{\lambda} P
 + \frac{d(d+2)}{4P^2} \, \partial_\kappa P \partial_\lambda P, \\*[2mm]
  P^{d/2} \,\partial_{\kappa} \partial_{\lambda} \partial_{\mu}
  \left( P^{-d/2} \right) & = &
  -\frac{d}{2P}\, \partial_{\kappa} \partial_{\lambda} \partial_{\mu}
                                P \nonumber \\*[2mm]
  & & +\frac{d (d+2)}{4P^2}\, [\partial_{\kappa}P \,
                     \partial_{\lambda} \partial_{\mu}P+
    \partial_\lambda\, P \partial_{\kappa} \partial_{\mu} P + \partial_\mu P
           \partial_{\kappa} \partial_{\lambda} P]  \nonumber  \\*[2mm]
        &  &  -\frac{d(d+2)(d+4)}{8P^3}\,\partial_\kappa P
              \partial_\lambda P \partial_\mu P.
\end{eqnarray}

This sequences of relations is general and holds for all graphs.
Each higher derivative follows from the previous one. The number of
terms grows since we have to calculate the derivatives of
a (negative) power of $P$. Each
term of the lower derivative yields a number of terms in precisely
the same way as the partitions of a set with $q$ elements follow
from those with $q-1$ elements.
So the systematics from the sequence is clear:
consider for the last Eq.~(\ref{c6}) all the partitions
of a set $\kappa,\lambda,\mu$ into groups and take the products
of these groups of derivatives.
The dimensional weights of these terms depend
only on the number of groups (or the number of factors). The sequence
also shows that the number of terms grows very rapidly with the number
of derivatives involved, as rapid as the number of
partitions of a set grows with the number of elements in the set.

The partitions are independent of the polynomials $P$ and they can
be generated once for all graphs and virial coefficients. The problem
is that the number increases so fast that memory problems arise
sooner or later for the higher number of edges. Therefore we have
to design strategies for handling this problem.

\section{Algorithms for the higher derivatives}

We outline here two complementary methods of evaluating the
coefficients $B_{l,k}$. The first uses the partitions and the
second computes the derivatives recursively. All numerical
calculations were carried out for dimension $d=3$.

\subsection{The method using partitions}

Here the basic ingredients are the derivatives of the Kirchhoff
polynomial $P$. Their values are integers: the number of terms that
survive after the differentiation. Since all $q$-th order
derivatives vanish for $q>=l$, there are not so many non-vanishing and
to begin with, they can be listed for each graph.

Then we have to carry out three summations:
\begin{enumerate}
\item The basic summation over the graphs qualifying for the virial
  coefficient.
\item For the coefficient $B_{l,k}$ all the choices of $k$ edges out
  of the set of edges of the graph.
\item For each choice of $k$ edges the summation over all partitions
  of the $k$ edges. Each partition contributes a product of the listed
  derivatives of $P$.
\end{enumerate}

This is straigthforward and fast, but gets
time comsuming and inaccurate for the highest derivatives. The errors
occur because the partitions get a sign depending on the number of
factors: even numbers of factors have a positive weight and odd numbers
a negative weight, while their contributions are often comparable.
These errors are avoided by integer arithmetic. Time and accuracy is
gained by using the symmetries of the graph.  Several choices of
$k$ edges can be equivalent due to the symmetry of the graph. E.g. any
choice of two edges in the case of the third virial coefficient gives
the same result. Calculating one of them and multiplying with the
number of equivalent choices is sufficient.

Without integer arithmetic the method works satisfactorily up to
11th derivative, with integer arithmetic we could extend that by
a few more.

In some cases the sum over the various derivatives simplifies due to
the following rule:
\begin{equation} \label{d1}
  \sum_\sigma \partial_\sigma P= {l-1 \brack q} P.
\end{equation}
Here $\sigma$ is a set of indices $(\kappa,\lambda, \cdots )$ with $q$
terms. The proof of this rule is based on the fact that one can select
in $l-1$ over $q$ ways an edge in each term of the spanning trees. All
these selections contribute 1 (after setting the $\alpha_i=1$) and the
total involves the number of terms $P$ in the spanning tree of the
graph. Doing the summation over the derivatives in Eqns.~(\ref{c6})
simplifies the outcome for to
\begin{equation} \label{d2}
  \sum_\sigma P^{d/2} \left( \partial_\sigma P^{-d/2} \right) =-\frac{d}{2}
  {l-1 \brack q}.
\end{equation} 
This rule can be used for the first term on the right hand
sides of the Eqns.~(\ref{c6}). For the first power of $A$ one finds
\begin{equation} \label{d3}
  B^*_{l,1} = - d (l-1) B^*_{l,0}.
\end{equation}
The proof of this relation follows from the property Eq.~(\ref{d2}). 

\subsection{The recursive method}
The second method computes the derivatives of $P^{-d/2}$ recursively.
Let again $\sigma$ be a choice $\kappa, \lambda, \cdots, \mu, \nu$
of $k$ edges of the graph. As the order of the differentiations does
not matter we order them in increasing such that $\nu$ is larger than
the preceding ones.
We can compute the derivative with respect to the last chosen
edge $\nu$ from the derivative of the predecessing choice
$\kappa, \lambda, \cdots, \mu$. These derivatives are ratios
of two polynomials and can be written as
\begin{equation} \label{d4}
  \partial_\sigma P^{-d/2} = \frac{R_k}{P^{d/2+k}}.
\end{equation}
The denominator is explicit and the numerator obeys the recursion
\begin{equation} \label{d5}
  R_{k+1} = P \times \partial_\nu R_k - (d/2+k)\, 
  \partial_\nu P \times  R_k.
\end{equation}
To make the recursion complete we set $R_0=1$.

To illustrate the
recursion we take the case $l=3$ as example with $P$ given by
Eq.~(\ref{c4}). The first level then gives 
\begin{equation} \label{d6}
  R_1 = P \times \partial_1 R_0 -(d/2) \partial_1 P \times R_0=
  -(d/2) (\alpha_2 + \alpha_3).
\end{equation}
The next level leads to 
\begin{equation} \label{d7}
  R_2 = P \times \partial_2 R_1 -(d/2+1) \partial_2 P \times R_1=
  - (1+2 \alpha_3) \times (d/2) +(d/2)(d/2+1) (1 + \alpha_3)^2.
\end{equation}
Here we encounter a new element. After a differentiation is carried
out, we may set the corresponding $\alpha=1$. So the factor $P$
in $R_2$ reduces to $P=1+2 \alpha_3$ and $R_2$ becomes a function
of the last remaining variable $\alpha_3$.
\begin{equation} \label{d8}
  R_2=(d/2)[(d/2)(1+2 \alpha_3+\alpha_3^2)+\alpha_3^2].
\end{equation} 
The final $R_3$ is a number
\begin{equation} \label{d9}
  R_3 =(d/2)(d+2)(2d-1).
\end{equation}
Note that this result agrees with the values given in Eq.~(\ref{c6}).

Characteristic for this method is that the polynomials increase in
length with the number of differentiation $k$ due to the
multiplications, but they shrink because the increasing number of
$\alpha$'s which can be put equal to 1. The latter property is
strengthend due to our choice to carry out the differentiations
in the order of increasing edge index. That means that all the
$\alpha$'s with index lower than the differentiated $\alpha$ may
be put equal to 1, since they will not occur in a further choice.
When all the $k$ differentiations are carried out 
the resulting polynomial is a number.

While the case, where three differentiations have to be carried out,
can be done by hand, higher orders have to be programmed. The
problem is to add and multiply polynomials. Therefore the terms
in the polynomial have to be coded. They have a coefficient in
front and a row of powers of the variables $\alpha$. The variables
$\alpha$ which are set equal to 1 get a power 0. Multiplying two
terms implies multiplication of the coefficients and addition of
of the exponents. Since the factors involving $P$ have at most a
power 1, in each multiplication the maximum power is raised by 1.
\begin{figure}[h]
\begin{center}
  \epsfxsize=0.7\linewidth  \epsffile{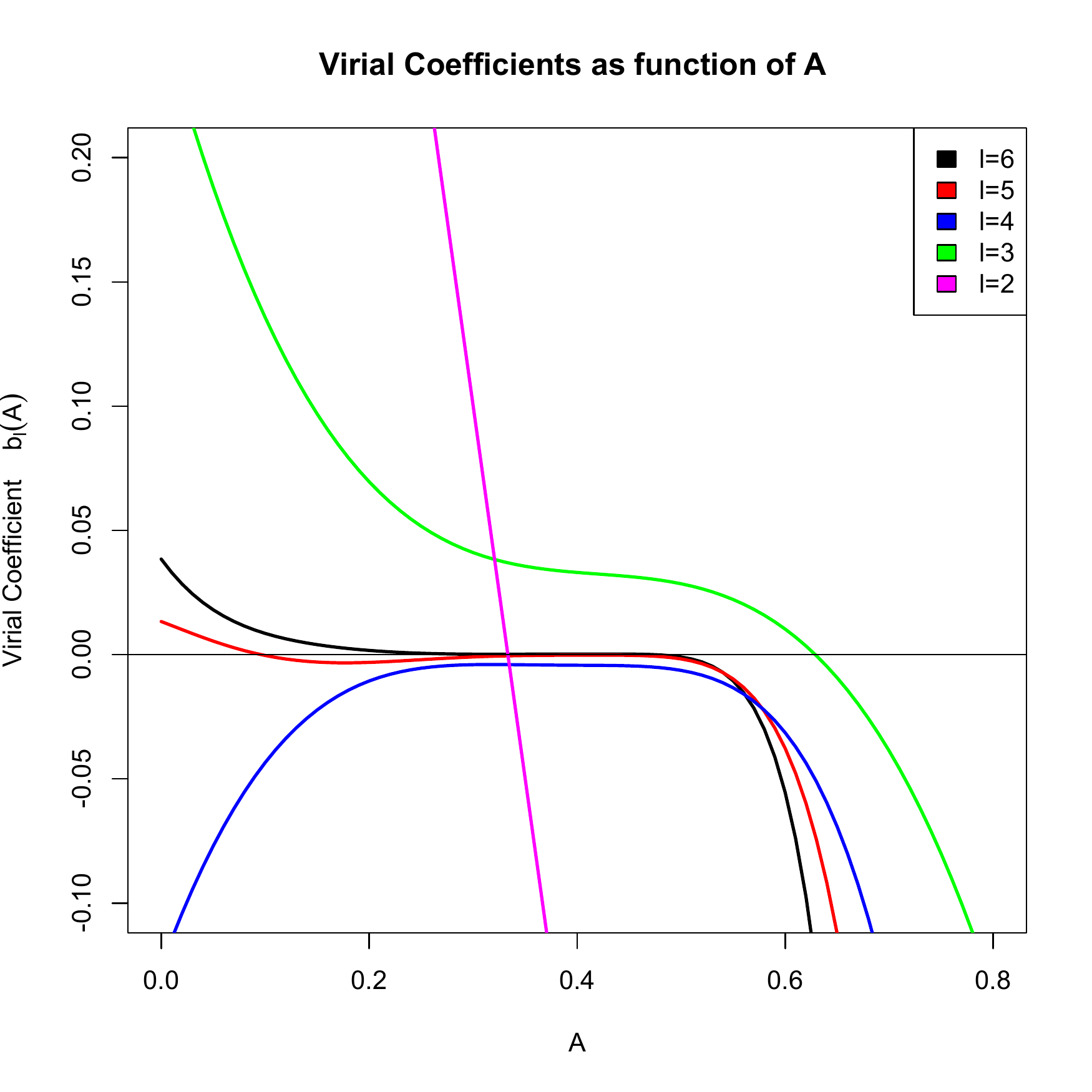}
  \vspace*{-0.1cm}
  
  \caption{Virial Coefficients $B_l^*(A)$}
  \label{coef}
  \end{center}
\end{figure}

It means that powers never exceeds $k$. (E.g. $R_2$ in Eq.~(\ref{d8})
has as highest power 2.) So powers form
a row of $k$ digits. The digits are limited by the order $k$ of the
differentiation. 
The factors
$P$ and $\partial_\nu P$ contain the $\alpha_i $ to atmost order 1.
Therefore the maximum power in $R$ increases in each recursion by 1.
After setting $\alpha_\nu=1$ the number of remaining $\alpha_i$
decreases by 1. So the expression for $R_k$ first increases in
complexity by increase of $k$ and finally reduces to a single
number when $k$ reaches the maximum value $m$ of edges.
In symmary the highest
power in $R_k$ is therefore $k$ and the number of remaining variables
is $m-k$. This makes the method efficient for the larger $k$. A graph
with $q$ edges contributes to the $B_{l,k}$ with $k \leq q$.

The two methods are
complementary: the first spends the most time in calculating the
highest derivatives, the second is slowest in the medium derivatives
and speads up towards the largest derivatives. The two have to give
the same result and this is a superb check on the calculation. For
$l=6$ the polynomial method spends too much time on the last few
graphs with 14  and 15 edges to complete the calculation, while the
method using the partitions just barely makes it in an acceptable
running time.

Below we have listed the coefficients for $l$ up to 6 and $k$ up to
10 for dimensions $d=3$.
\begin{center}
  \begin{tabular}{|c|r|r|r|r|r|r|}
    \hline 
    & & & & & & \\[-2mm]
    $k \backslash l$ &$ l=2$ \hspace{3mm} &  $l=3$ \hspace{3mm} & $l=4$         \hspace{3mm}
    & $l=5$ \hspace{3mm} &    $l=6$  \hspace{3mm} & $l=6$ approx.\\[2mm]
    \hline 
    & & & & & & \\[-2mm]
    $k=0$ & 1.0000 & 0.25660 & -0.12546 & 0.01333 & 0.03846 &0.03846 \\[2mm]
    \hline
    & & & & & &\\[-2mm]
    $k=1$ & -3.0000 & -1.53960 & 1.12914 & -0.15990 & -0.57691 &
                                            -0.57691    \\[2mm]
    \hline
    & & & & & & \\[-2mm]
    $k=2$ & &3.59240 & -3.31095  & -0.23737 & 4.11313 & 4.11313 \\[2mm]
    \hline 
    & & & & & & \\[-2mm]
    $k=3$ & & -2.85111 & 1.55525 & 6.69183 & -16.29557 & -16.29557 \\[2mm]
    \hline 
    & & & & & & \\[-2mm]
    $k=4$ & & & 7.33549 &-21.76907& 28.93441 & 28.93441 \\ [2mm]
    \hline 
    & & & & & & \\[-2mm]
    $k=5$ & & &  -8.35278 & 15.29319 & 15.61594 & 15.68595\\[2mm]
    \hline 
    & & & & & & \\[-2mm]
    $k=6$ & & &  -0.61340 & 28.93520 & -130.36214 & -130.02966 \\[2mm]
    \hline 
    & & & & & & \\[-2mm]
    $k=7$ & & & & -32.09542 & 91.70453 & 88.92666 \\[2mm]
    \hline 
    & & & & & & \\[-2mm]
    $k=8$ & & & & -7.51715 & 156.06942 & 150.41702 \\[2mm]
    \hline 
    & & & & & & \\[-2mm]
    $k=9$  & & &  & -0.79677 & -130.51623 & -134.80666 \\[2mm]
    \hline
    & & & & & & \\[-2mm]
    $k=10$ & & &  & -0.04424 & -65.6438 & -66.94143\\[2mm]
    \hline
  \end{tabular}
\begin{minipage}{14cm}
  Table 2.  Polynomial coefficients $B_{l,k}$ for the extended
  UF-model.
  The 5 further coefficients $B^*_{6,k}$ for $k=11,12,13,14$ and
$k=15$ are respectively: \mbox{-16.65300,} -2.69820, -0.59634, 0.14819
and -0.04533.
The last collumn refers to the approximation described  in the Appendix.
\end{minipage}
\end{center}
In order to see what these polynomial coefficients mean for the
virial coefficient $B^*_l (A)$ we plot these coefficiens as function of
$A$ in Fig.~(\ref{coef}). Apart from $B_2^*(A)$, which depends
linearly on $A$, they drop off rapidly to small values in the range
$A=0.2$ to $A=0.6$. Beyond $A=0.6$ they evolve to larger negative
values. The range $A=0.2$ to $A=0.6$ of small virial coefficients
point to a delicate interplay of the terms is the polynomial expression
for $B_l^*(A)$. So the $B^*_{l,k}$ have to be computed with high
precision. In order to see how small the values become near A=0.5,
we list here the values of $B_l^*(A_c)$ with
$A_c=0.4259$ the critical value (see next Section).
$B_2^*(A_c)=-0.277700, B_3^*(A_c)=0.032252, B_4^*(A_c)=-0.004338,
B_5^*(A_c)=-0.002205, B_6^*(A_c)=0.000243$.

The last column of Table 2, gives the values of $B^*_{l,k}$
following from the approximation described in the Appendix.

\section{Phase Diagram}\label{phasedia}

Speculating on the phase diagram of the extended UF model, we note
that the potential does not have a hard core with a finite range
or a sharp deep well,
which may induce the geometric ordering in a crystal.
So we do not expect
that the model has a transition to an ordered solid phase at high
densities.

On the other hand the attraction may lead to a gas-fluid phase
transition at intermediate densities. As a signal of this transition
one would see the appearance of a van der Waals loop in the pressure.
Such a loop is the result of an interplay of a negative second
virial coefficient bending the pressure down and positive higher virial coefficients which
turn the pressure upwards again at higher densities.

We note that in Table 2 the $B_{l,m}$, with $m=l(l-1)/2$ the highest
power of $A$, all are negative. That means that for large $A$ the
virial coefficients (beyond the ideal gas term) become negative
and that there is no stability in the pressure at high pressures.
At intermediate
values of $A$, $0.4 < A < 0.6$, the highest $B_{l,m}$ is positive
for $l=3$ and $l=6$. For these cases a van der Waals loop
may occur.

\begin{figure}[h]
\begin{center}
  \epsfxsize=0.7\linewidth  \epsffile{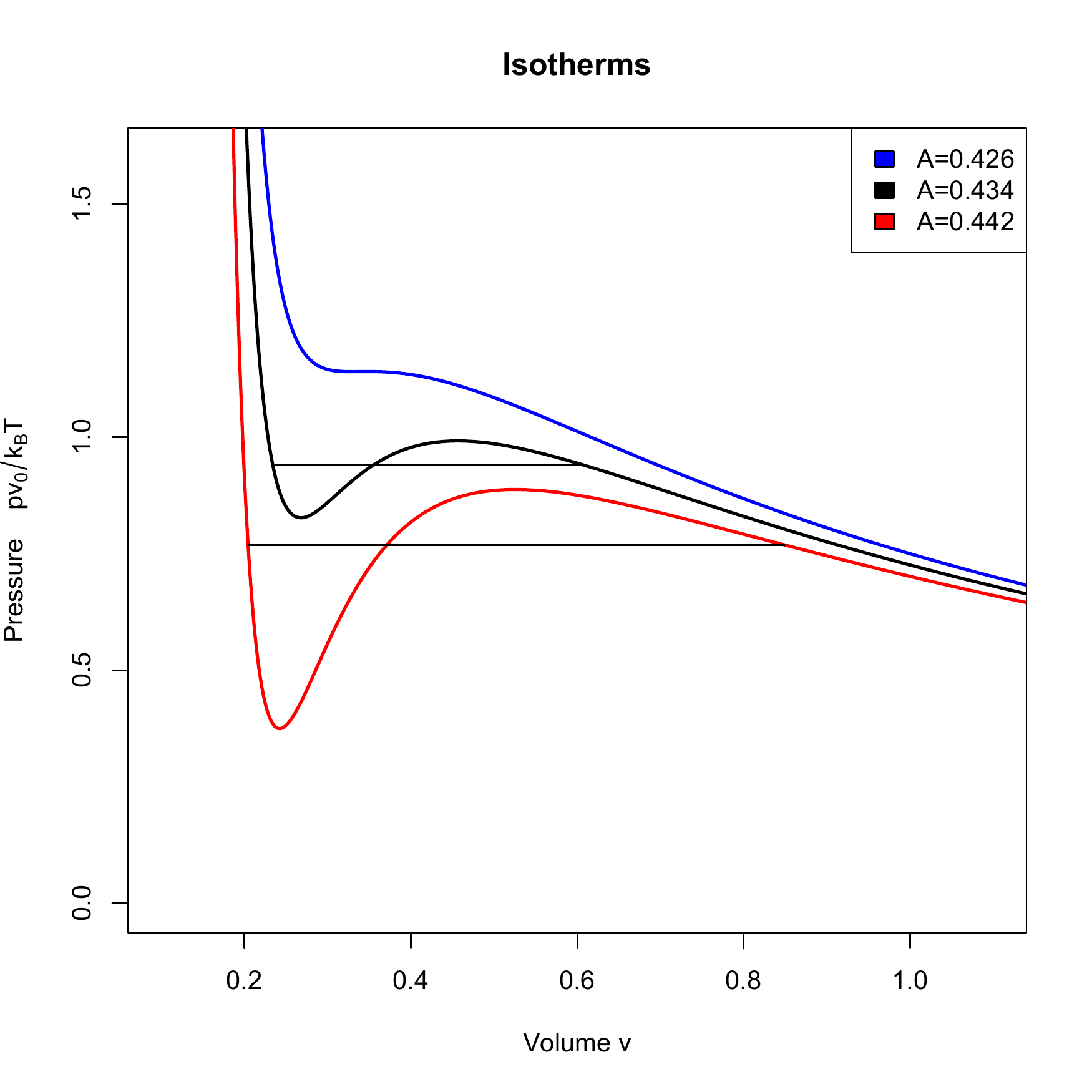}
  \vspace*{-0.1cm}
  
  \caption{Critical and subcritical ``isotherms''. The tielines give
  coexsisting phases.}
  \label{pressure}
  \end{center}
\end{figure}

The onset of the loop, the critical point, is found from the conditions
\begin{equation} \label{x1}
  \frac{\partial p}{\partial n} =0, \quad \quad \quad
  \frac{\partial^2 p}{\partial n^2} =0.
\end{equation} 
Using the virial series for the pressure, the equations get the form
\begin{equation} \label{x2}
  \sum_l l B^*_l(A) \, n^{*(l-1)} =0;
\end{equation}
and
\begin{equation} \label{x3}
  \sum_l l (l-1) B^*_l(A) \, n^{*(l-2)} =0;
\end{equation}
From Eq.(\ref{x3}) one sees that at least three terms are needed for
a solution. In that case the equation becomes a linear equation in
$n^*$. Plugging the solution for $n^*$ into Eq.~(\ref{x2}) one finds
the condition for $A$
\begin{equation} \label{d10}
  [B^*_2(A)]^2 =3 B^*_3(A)
\end{equation} 
Using Eq.~(\ref{c2}) for $B^*_2(A)$ and Eq.~(\ref{c6}) for $B^*_3(A)$
gives a cubic equation for $A$, with the solution $A_c=0.43647$.
For the virial series terminating at $l=6$  one has to solve
the equation numerically, with result $A_c=0.42591$
In Fig.~(\ref{pressure}) we show a few ``isothermes'',
the critical and two subcritical exhibiting a vander Waals loop, with
the equal area construction for the coexisting phases.
  
This gives an indication of the gas-fluid transition and its location
in the phase space. Of course one
cannot derive the critical singularities from a finite virial
series. It is likely that the critical point of the model with
attraction is in the class of the 3d-Ising model.

\section{A Connection to standard potentials}\label{lennard}

In order to get an impression of the length scale $a$ and the
attraction amplitude $A$, we compare the a few virial coefficients
of the model, with those of $2n-n$ Lennard-Jones potentials. For $n=6$
this potential is often used
for simple molecules like the noble gases. They are of the form
\cite{main}
\begin{equation} \label{f1}
  V(r) = 4 \epsilon \left[ \left(\frac{\sigma}{r}\right)^{2n} -
    \left(\frac{\sigma}{r}\right)^{n} \right].
\end{equation}
$\epsilon$ is the depth of the potential well and $\sigma$ the range
of the interaction. $\epsilon$ can be used for scaling  the
temperature $T$ to the dimensionless $T^*$ as
\begin{equation} \label{f2}
  T^* = k_{B} T/\epsilon.
\end{equation}
Let $f(r)$ be the associated Mayer function according to
Eq.~(\ref{a3}), then the second virial coefficient $B_2$ can be
written as
\begin{equation} \label{f3}
  B_2 = \sigma^3 \, B(T^*) = - \frac{4 \pi}{2} \int r^2dr f(r).
\end{equation} 
$B(T^*)$ is a dimensionless function of $T^*$. 

Comparing this with the expression Eq.~(\ref{d1}) for the second
virial coefficient we see that one can tune the parameter $a/\sigma$
such that the two values are equal with one proviso: the parameter
$A$ has to be chosen such that both virial coefficients have the
same sign. That means that for higher $T^*$, where the Lennard-Jones
second virial coefficient is positive, $A<1/3$ and that for the lower
temperatures $A>1/3$. With this restriction $A$ is a still a free
parameter which can be tuned such that also the third coefficients
of the representations coincide.

The third virial coefficient is given by
\begin{equation} \label{f4}
  B_3= \sigma^6 C(T^*) = - \frac{1}{3} \int d {\bf r_2}
  d {\bf r_3} f(r_{12}) f(r_{23}) f(r_{31}).
\end{equation}
Using the fourier transform
\begin{equation} \label{f5} 
  \tilde{f} (k) = \frac{1}{(2 \pi)^3 } \int d \bf{r} \, f(r)
  \exp (i \bf{k \cdot r}),
\end{equation}
the third virial coefficient reads
\begin{equation} \label{f6}
  B_3 = - \frac{4 \pi}{3 (2 \pi)^3} \int k^2 dk 
  [\tilde{f} (k)]^3.
\end{equation}
The ratio
\begin{equation} \label{f7}
  \frac{B_3}{(B_2)^2} = \frac{C(T^*)}{B^2(T^*)},
\end{equation}
is independent of the scale $\sigma$ and may be used for matching
the third virial coefficient.

\begin{figure}[h]
\begin{center}
  \epsfxsize=\linewidth  \epsffile{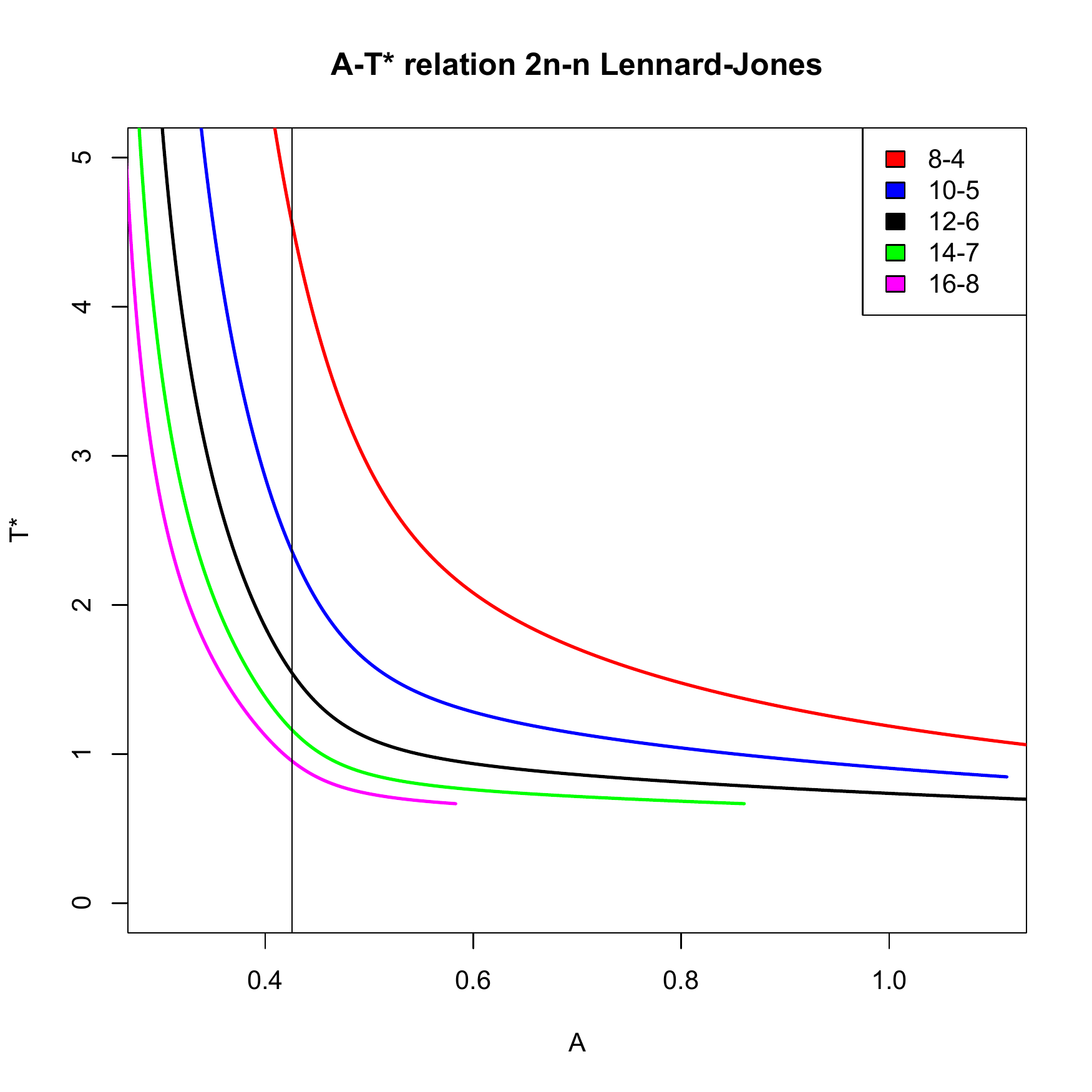}
  \vspace*{-8mm}
  
  \caption{The relation between of $A$ and $T^*$, for a number of
    2n-n Lennard-Jones potentials. The vertical line corresponds to
  the critical $A_c=0.4259$.}
\label{mayer}
\end{center}
\end{figure}

The corresponding ratio for the soft potential is contained in
Eq.~(\ref{c4}) and reads explicitly
\begin{equation} \label{f8}
  \frac{C(T^*)}{B^2(T^*)} =\frac{4}{9 \sqrt{3}}\frac{1-6 A + 14 A^2 -
  (100/9) A^3 }{(1-3 A)^2}.
\end{equation} 
This gives a relation between $T^*$ and $A$. Note that this relation
does not involve the ranges of the two models which are compared.
In Fig.~(\ref{mayer}) we
show for a few choices of $n$ the value of $A$ that corresponds to
$T^*$. This relation implies an estimate for the critical temperature
$T^*_c(n)$. We find $T^*_c(8)=0.95, T^*_c(7)=1.16, T^*_c(6)=1.54,
T^*_c(5)=2.36, T^*_c(4)=4.56$, which are reasonable estimates, given the
crudeness of the match and which show the proper trend. The vertical
line corresponds to the critical value $A_c=0.4259$ 

\section{Conclusion}

We have extended the Uhlenbeck-Ford model with an attractive part,
having an amplitude $A$. The corresponding potential is drawn for
$A=1$ in Fig.~\ref{potential}. Characteristic is the soft repulsive
core and the rather extended attractive well. The virial coefficients
$B_l$ become functions $B_l(A)$ in the form of a finite
power series in $A$, with highest power $m=l(l-1)/2$.
The calculation of the coefficients
of these power series is considerably more involved than that of
$B_l$ for $A=0$ (pure repulsion). We have determined $B_l(A)$
for $l$ up to 6.

The key quantity in the calculation is the Kirchhoff polynomial
$P(\alpha_i)$, defined in Section \ref{gauss}.
It is a polynomial in the variables
$\alpha_i$, based on the spanning trees of the graph. Each of the
$l-1$ edges of a spanning tree contributes a factor
$\alpha_i$ if the $i$-th edge is present in the graph.
For $B_l(A)$ one needs the derivatives of $P^{-d/2}$.
We have employed two methods: one using the partitions of the
$\alpha_i$ and one using recursively the derivates of the function
$P(\alpha_i)$. The mutual agreement of these independent methods is
a guarantee of the correctness of the coefficients.

One may expect that with stronger attraction the model forms
a fluid phase. Indeed for intermediate $A$ we observe the formation
of a van der Waals loop with a critical value $A_c=0.246$. An
intruiging question is the behaviour for large values of $A$ of the
functions $B_l(A)$. The sofar calculated highest coefficients
$B_{l,m}$ become negative, which implies that the functions $B_l(A)$
becomes negative for large $A$. That leads to an instability in
the virial series through negative values  of the pressure.

We have made a link between the amplitude $A$ of the attraction and
$1/T^*$ in Lennard-Jones type interaction. 

An approximation
scheme has been developed (see the Appendix) which works well for the
lower derivatives.

We close with the remark that the graph contributions to the pair
correlation function are also exactly calculable as a power series
in the density, but this is another project, more complicated than
the present one.

{\bf Acknowledgement}. The author is indebted to Bernard Nienhuis
for critical comments on the manuscript and 
for his advice on the mathematical aspects of graph theory and to
Guus Regts for making graph counting routines available.
He also thanks
Marc van Leeuwen for providing a library handling big integers and
Henk Lekkerkerker for stimulating discussions.

\appendix
\section{Approximation}\label{approx}
As mentioned the number of partitions of a set increases very rapidly
with the number of elements (edges of the graph). Therefore the
computation of the virial
coefficients spends most of the time calculating the contribution of
the (almost) fully occupied graphs. We also
observed that the various derivatives of a set of edges vary
little with the specific edges in the set. Only the size $q$ of the
set is the main ingredient. Note that for the average value one can
use the rule Eq.~(\ref{d2}).

Based on this idea we have designed an fast scheme, which give
an indication of the magnitude of the contributing terms.
Replacing the individual derivatives
by the average over a set of $q$ edges, many partitions give the same
value for the derivative. Thus we can lump these partitions together.
In fact all the permutations of the edges which keep the sizes of the
bins equal, yield the same contribution. Their number is the
multinomial
\begin{equation} \label{e11}
  N(\{\sigma\}) = \frac{\sigma!}{(\sigma_1! \sigma_2! \cdots)(p_1!
    p_2! \cdots)}.
\end{equation} 
Here $\{\sigma\} = \sigma_1, \sigma_2 \cdots $ is a set of bin sizes with
$\sigma_1+\sigma_2 \cdots =\sigma$ and $p_1$ the number of bins of
size 1, $p_2$ of size 2 $\cdots$, with $p_1 + 2 p_2 + \cdots
=\sigma$.
The contribution of the selection of partitions then equals
\begin{equation} \label{e12}
  D(\{\sigma\}) = N(\{\sigma\}) \prod w_{\sigma_1} w_{\sigma_2} \cdots
\end{equation}
This is an enormous reduction in the number of partitions. For
instance for $k=15$ edges to be distributed over 4 bins, one has
42355950 detailed partitions which all lead to the same average
value. In fact one needs only the partitions of the edges over bins,
no matter which edge is in which bin (the partitions of a number).

If one wants to do better than this lowest approximation, one must
take the fluctuations into account. This can be done by giving each
bin of edges the sum of the average value of the derivative and the
deviation thereof (the fluctuation). The lowest approximation is
taking only average values. The first approximation is taking one
fluctuating bin combined with averages of the others (yields a
vanishing contribution), the second by
taking two fluctuating bins combined with averages etc. So one still
needs the full partitions of the fluctuating bins, but that is
limited as long as their number is small. This gives a quick estimate
of the value of $B^*_{l,k}$ provided that $k$ is not too large.
In fact if one takes into account the products of fluctuation to order
$n$ the first cofficients $B^*_{l,k}$ with $k \leq n$ becomes exact.
In Table 2, last column, we have given the approximate values, taking
the fluctuations into account to 4th order. For all other values in
the Table the approximation gives virtually the exact value. Beyond
$k=10$ and $l=6$ the approximation gets inaccurate, but these
coefficients are less important for $A<1$.

As Table 2 shows the approximated $B^*_{6,k}$ are quite close to the
exact values. But these small differences can build up to important
differences in the value of $B^*_l (A)$, as the individual terms
$B^*_{6,k} A^k$ are much larger than the sum $B^*_l (A)$
for $A \simeq 0.5$, which is is an interesting value of $A$
(see next Section (\ref{phasedia}).

\end{document}